\documentstyle[prd,aps,eqsecnum]{revtex}
\newcommand{\bmit}[1]{\mbox{\boldmath $#1$}}

\newcommand{\gm}{\left[\frac{k_n}{\sqrt{2}}\,\left(x^1+x^3-
\frac{H\,f^{(-1)}}{2}\right)\right]}
\newcommand{\alf}{\left(k_n x^0 + \frac{k_n\,H\,f^{(-1)}}{2}\right)}
\newcommand{\nb}{{\bar{n}}}
\begin{document}

\draft
%\twocolumn
%@twocolumnfalse\endcsname
\preprint{Submitted to Class. Quant. Grav.}
\title{Exact Solution to the homogeneous Maxwell Equations in the 
Field of a Gravitational Wave in Linearized Theory}
\author{Mirco Calura and Enrico Montanari\footnote{Electronic address: 
montanari@fe.infn.it}}
\address{Department of Physics, University of Ferrara and
INFN Sezione di Ferrara, Via Paradiso 12,
I-44100 Ferrara, Italy}
%\date{\today}
\maketitle
\begin{abstract}
We present the exact solution to the linearized Maxwell equations in 
space--time slightly curved by a gravitational wave. We show that in 
general, even dealing with a first--order theory in the strength of 
the gravitational field, the solution can not be written as the 
sum of the flat space--time one and a weak perturbation due to the 
external field. 
%It is shown 
%We show that such 
Such an impossibility arises when either the frequency of the 
gravitational wave is too low or too high with respect to the 
one of the electromagnetic field. 
We also provide an application of the solution to the case of 
an electromagnetic field bounced between two parallel conducting 
planes.
%a resonant cavity 
%is provided. 

\end{abstract}

\section{Introduction}

The problem on the propagation of electromagnetic 
fields in a curved space--time has been widely studied all through the 
years. 
Einstein himself calculated the deflection of a light ray by the 
gravitational field of the Sun.
This problem can be simply approached by assuming light 
rays as trajectories of a zero--mass particle, 
for which $ds^2 = 0$. Moreover the same 
approach has been used to calculate the response of an interferometer 
to an impinging gravitational wave~\cite{saul}. 

A different approach to the problem is considering directly 
the Maxwell equations 
in curved space--time. Two kinds of phenomena 
can be described by this method: propagation of electromagnetic 
radiation in a non--radiative gravitational 
background~\cite{bal58,ple60,hee64,pos65,muh66} 
(see also Ref.~\cite{sch92} for an 
exhaustive treatise on the subject) and in the field of a 
gravitational wave. Not only are they interesting problems themselves 
but also have important experimental applications (see for 
instance~\cite{alc96}).
 
As for the propagation in a gravitational wave field, some authors 
have solved the problem by using the formalism of geometrical optics; 
therefore these results are valid 
under the assumption that the 
gravitational wave is slowly varying with respect to the 
electromagnetic field \cite{zip66,lob92,mon98}. 
Others have solved the equations by splitting the 
electromagnetic tensor (or the four--vector potential) in a sum of two
terms: the unperturbed solution and a correction 
term~\cite{coo68,boc70,ber71,bra73,fun78,cod80,bar85,bra90,coo93}. 
However, as it has been shown in Ref.~\cite{mon98}, the splitting 
approach turns the problem of the propagation of a free 
electromagnetic field in a gravitational background into a different 
one, that is the generation of a field by a given current. On this 
ground many doubts could be raised as to the validity of this 
procedure (a detailed discussion may be found in Ref.~\cite{mon98}).
%Hence 
%, at least in principle, 
%this procedure is not a suitable one.

Therefore we see that a satisfactory solution that only 
assumes the smallness of the gravitational wave amplitude regardless 
of its shape or frequency is lacking.  
In this paper we provide exact solution to the problem by solving 
%the 
Maxwell equations for the 4--potential (De Rham equations) 
in the framework of the linearized general relativity (Sec. II). 
In other words the solution we have achieved is correct as long as the 
linearized De Rham equations~(\ref{derahml}) describe properly the 
physical problem. We express our result as an 
integral which is the curved space--time version of the usual Fourier 
integral. In this way any solution can be obtained by a proper 
choice of the Fourier--like coefficients.
Afterwards we calculate the expressions of the electromagnetic 
tensor components (Sec. III). Finally Sec. IV is devoted to the 
application of our solution to a simple case in order to 
compare it to the one obtained through the ``splitting'' 
procedure. 

Besides we find that the main effect of a low--frequency 
gravitational wave on 
an electromagnetic field bounced between two parallel conducting walls
%a resonant cavity 
is ohmic power loss. 
%in their bodies.
%the body of its conducting walls.

\section{Exact solution to the de Maxwell equations}

The homogeneous de Rham equations describe the propagation of 
electromagnetic field in an external curved space--time. Under the 
Lorentz condition, they could be written as:
\begin{equation}
\left \{ \begin{array}{l}
A^{\mu;\alpha}_{\ \ ;\alpha} 
%- R^{\mu}_{\ \nu} A^{\nu} 
= 0
%\\ \ 
\\
A^{\sigma}_{\ ;\sigma} = 0
\end{array} 
\right .
\label{derahm}
\end{equation}
(throughout this paper the notation and conventions are as 
in~\cite{mtw} exept for indices $p$, $r$, $s$ which take values 
1 and 2).
Let us assume that the curvature of space--time is due to a weak 
plane gravitational wave, propagating in the $x^3$ direction of a 
given {\em transverse traceless} (TT) reference frame~\cite{mtw}.
In this case the metric tensor is:
\begin{equation} 
g_{\mu\nu} = \eta_{\mu\nu} + h_{\mu\nu}(x^3-x^0),\qquad\qquad
|h_{\mu\nu}| \ll 1,
\label{lincond}
\end{equation}
where $\eta_{\mu\nu}$ is the Minkowski metric tensor with positive 
signature and
$h_{\mu\nu}(x^3-x^0)$ is the perturbation to the flat space--time 
metric. In the chosen reference frame the only non--vanishing 
components are $h_{11} = - h_{22} = h_+$ and 
$h_{12} = h_{21} = h_\times$.
Neglecting 
second order terms 
%terms of second order 
in $h_{\mu\nu}$, Eqs.~(\ref{derahm}) 
become ($h^\mu_{\ \nu} = \eta^{\mu\alpha}\,h_{\alpha\nu}$):
\begin{equation}
\left \{ \begin{array}{l}
A^{\mu,\nu}_{\ \ \,\nu} - h^{\alpha\beta} A^\mu_{\ ,\alpha,\beta} +
\left (
h^{\mu\ ,\beta}_{\ \alpha} + h^{\mu\beta}_{\ \ ,\alpha} -
h_\alpha^{\ \beta,\mu} \right ) A^\alpha_{\ ,\beta} = 0 \\
%\ \\
A^\sigma_{\ ,\sigma} = 0
\end{array}
\right.
\label{derahml}
\end{equation}
Since the first equation of the above system for $\mu=0$ is the same 
as $\mu=3$, we can choose a particular Lorentz gauge in which 
$A^3 = A^0$. This greatly simplifies the problem, because this 
way the equations are partially decoupled [see 
Eqs.~(\ref{tcomp})--(\ref{coulomb}) later on].
Eqs.~(\ref{derahml}) are a set of differential equations with variable 
coefficients, which depend only on $x^3-x^0$. This fact leads us to perform 
the following coordinate transformation:
\begin{equation}
\left \{ \begin{array}{rcl}
X^0 & = & (x^3-x^0)/\sqrt{2} \\
X^1 & = & x^1 \\
X^2 & = & x^2 \\
X^3 & = & (x^3+x^0)/\sqrt{2}
\end{array} \right.
\label{uv}
\end{equation}
With this new set of variables, system~(\ref{derahml}) becomes
($\partial_\mu$ means $\frac{\partial\ \,}{\partial X^\mu}$):
\begin{eqnarray}
\bmit{\hat D}{\cal A}^p +
\partial_0 h^p_{\ s} \partial_3 {\cal A}^s = 0 \label{tcomp} \\
\bmit{\hat D}{\cal A}^3 -
\partial_0 h_r^{\ s} \partial_s {\cal A}^r = 0 \label{vcomp} \\
\bmit{\hat D}{\cal A}^0 = 0 \label{ucomp} \\
{\cal A}^0 = 0; \qquad \qquad 
\partial_1 {\cal A}^1 + \partial_2 {\cal A}^2 +
\partial_3 {\cal A}^3 = 0.  \label{coulomb} 
\end{eqnarray}
where 
\[
\bmit{\hat D} = \left (\partial_1\partial_1 + \partial_2\partial_2 + 
2 \partial_3\partial_0 - h^{rs} \partial_r\partial_s\right ), 
\]
$p,r,s=1,2$, and 
${\cal A}^\mu=\frac{\partial X^\mu}{\partial x^\nu}\,A^\nu$ 
are the new components of the 
4--vector potential. We observe that in this new reference frame 
the electromagnetic gauge condition is the so--called Coulomb 
gauge~\cite{mon98}. Hereinafter only the real part of equations is to 
be retained. 
First of all we solve Eq.~(\ref{tcomp}); through Fourier 
transform in $X^j$ ($\bmit X \rightarrow (X^1,X^2,X^3)$, 
$\bmit\lambda \rightarrow (\lambda_1,\lambda_2,\lambda_3)$)
\begin{equation}
{\cal A}^p(\bmit X,X^0) = \int 
e^{i\,\lambda_k X^k}\, a^p(\bmit\lambda,X^0)\, d^3 \bmit\lambda,
\label{fourier}
\end{equation}
one gets:
\begin{equation}
\left[2i \lambda_3 \frac{d\ }{dX^0} - 
\left(\lambda_1^2+\lambda_2^2\right) + 
\lambda_r \lambda_s 
h^{r\,s}\left(\sqrt{2}\,X^0\right)\right]
\,a^{p} + 
%\sqrt{2}\,
i\,\lambda_3 \frac{dh_{r}^{\ p}}{dX^0}\,a^{r} = 0.
\label{eqp}
\end{equation}
Setting:
\begin{equation}
a^{p} = \gamma(X^0;\,\bmit\lambda)\,
B^{p}(\sqrt{2}\,X^0;\,\bmit\lambda)
\label{ansatz}
\end{equation}
with
\begin{equation}
%\gamma(X^0;\,\bmit\lambda) = \exp{\left\{\frac{1}{2i \lambda_3}
%\,\left[ (\lambda_1^2+\lambda_2^2) X^0 - 
%\lambda_r \lambda_s h^{(-1) r\,s}(\sqrt{2}\,X^0)\right]\right\}},
\gamma(X^0;\,\bmit\lambda) = \exp{\left[
\frac{(\lambda_1^2+\lambda_2^2) X^0 - 
\lambda_r \lambda_s h^{(-1) r\,s}(\sqrt{2}\,X^0)}
{2i \lambda_3}\right]},
\label{gamma}
\end{equation}
($h^{(-1) r\,s}$ stands for any primitive of $h^{r\,s}$)
Eq.~(\ref{eqp}) gives:
\begin{equation}
\frac{dB^p}{dX^0} + 
%\frac{1}{\sqrt{2}}
\frac{1}{2}\,\frac{dh^{p}_{\ q}}{dX^0}B^{q} = 0.
\label{eqB}
\end{equation}
The solution to the above equation could be expressed as a function 
series:
\begin{equation}
\begin{array}{rcl}
B^r(w;\bmit\lambda) & = & \left [ \delta^r_{\ p} +
\sum_{n=1}^{\infty} \frac{(-1)^n}{2^n}\ 
\aleph^r_{\ p}(n;w) \right ] b^p (\bmit\lambda) \\
\aleph^r_{\ p}(n;w) & = & 
\int \frac{d h^r_{\ q}}{dw}\ \aleph^q_{\ p}(n-1;w)\ dw \\
\aleph^r_{\ p}(1;w) & = &
\int \frac{d h^r_{\ p}}{dw}\ dw
\end{array}
\label{solB}
\end{equation}
where $b^p (\bmit\lambda)$ depends only on $\bmit\lambda$.
As one can easily see, $\aleph^r_{\ p}(n;w)$ are of 
same order as $h^n$, where $h$ is the order of magnitude of the 
gravitational wave amplitude. 
A sufficient condition for the series to converge is that
$\aleph^r_{\ p}(n;w) \leq 1$; this is always satisfied under the 
assumption of linearized theory, in which $h \ll 1$. Therefore, 
through this approximation, we can keep only the first term in 
the series~(\ref{solB}). In this case
\begin{equation}
B^r\left(\sqrt{2}\,X^0;\bmit\lambda\right) = 
b^r (\bmit\lambda) -\frac{1}{2} h^r_{\ p}\left(\sqrt{2}\,X^0\right)
b^p (\bmit\lambda).
\label{solBl}
\end{equation}

As for the ${\cal A}^3$ component, the
Coulomb gauge [Eq.~(\ref{coulomb})] gives:
\begin{equation}
a^3 = - \frac{\lambda_1\,a^1 + \lambda_2\,a^2}{\lambda_3} 
\label{sol3}
\end{equation}
which automatically fulfills Eq.~(\ref{vcomp}).

Having solved the system of 
equations~(\ref{tcomp})--(\ref{coulomb}) in the $X^\mu$ reference 
frame [Eqs.~(\ref{uv})], we proceed to write the solution in the 
$x^\alpha$ coordinates. In the case of a flat space--time, i.e. 
$h_{\mu\nu} = 0$, $k_j$ would be the Fourier variables conjugated to 
$x^j$ coordinates.
In this way we have that 
$\lambda_3=(k_3 \pm k)/\sqrt{2}$ ($k=\sqrt{k_1^2 + k_2^2 +k_3^2}$).
One gets ($\bmit k \rightarrow (k_1,k_2,k_3)$):
\begin{eqnarray}
A^p &=& \int{B^p(x^3-x^0;\bmit{k})\,e^{i\,\bmit{k}\cdot\bmit{x}}
\ \left[ \beta^+(\bmit{k})\,e^{i\,(kt + \psi^+)} +
\beta^-(\bmit{k})\,e^{- i\,(kt - \psi^-)}\right] 
d^3\bmit k}
\label{ap} \\
A^3 = A^0 &=& 
\int{k_s\,B^s(x^3-x^0;\bmit{k})\,e^{i\,\bmit{k}\cdot\bmit{x}}\ 
\left[ -\frac{\beta^+(\bmit{k})}{k+k_3}
\,e^{i\,(kt + \psi^+)} +
\frac{\beta^-(\bmit{k})}{k-k_3}\,e^{- i\,(kt - \psi^-)}\right]
d^3 \bmit k}
\label{a30}
\end{eqnarray}
where
\begin{equation}
\psi^{\pm} = \pm \frac{k_r k_s h^{(-1) r\,s}(x^3-x^0)}{2\,\left(
k \pm k_3 \right)}
\label{psi}
\end{equation}
and $\beta^\pm (\bmit{k})$ are generic functions whose actual form is 
determined by assigning suitable initial or boundary conditions. 
The solution given in Eqs.~(\ref{ap}) and~(\ref{a30}) only assumes  
the smallness of $h_{\mu\nu}$ and does not require any further hypothesis 
about its rate of variation. 
From Eq.~(\ref{psi}) it follows that:
\[
\psi \sim \frac{k}{\chi}\ h
\]
where $\chi$ is the order of magnitude for a typical frequency of 
gravitational wave. 
As no assumption was made over $\chi$,
$\psi$ is not necessarily a first order term in $h$. Therefore, in 
the general case, it is not possible to expand 
the exponential terms in the solution 
to the lowest order in 
$\psi$, opposite to the case of
the function series~(\ref{solB}), in which there is no term
proportional to $1/\chi$.

If the gravitational wave is linearly polarized then:
\begin{equation}
h_+(x^3-x^0) = H_+\,f(x^3 - x^0),\qquad\qquad
h_\times(x^3-x^0) = H_\times\,f(x^3-x^0)
\label{hlin}
\end{equation}
where $H_+$ and $H_\times$ are constants. We can set:
\begin{equation}
H_+ = H\,\cos{2 \alpha},\qquad\qquad
H_\times = H\,\sin{2 \alpha}.
\label{halpha}
\end{equation}
In this case the solution to Eq.~(\ref{eqB}) greatly simplifies. 
Actually it is possible to express it as a function rather than an 
infinite series of functions [see Eq.~(\ref{solB})]. 
Using standard methods the solution can be written:
\begin{equation}
B^p = b^p\,e^{- \frac{\lambda}{2}\,f(x^3-x^0)}
\end{equation}
where $b^p$ and $\lambda$ are the eigenvectors and eigenvalues of the 
following system:
\begin{equation}
H^{p}_{\ q}\,b^{q} = \lambda\,b^{p}.
\end{equation}
The eigenvalues are $\lambda=H$ and $\lambda=-H$ while the 
corresponding eigenvectors are:
\begin{equation}
\bmit b_+ = (\cos{\alpha},\sin{\alpha}) \qquad {\rm and}
\qquad
\bmit b_- = (\sin{\alpha},-\cos{\alpha})
\label{bpiumeno}
\end{equation}
respectively. Therefore the solutions read
\begin{equation}
B^p_+ = b^p_+\,e^{- \frac{H}{2}\,f(u)} 
\qquad {\rm and} \qquad
B^p_- = b^p_-\,e^{+ \frac{H}{2}\,f(u)}.
\end{equation}

\section{the electromagnetic tensor}

Having found the solution to the de Rham equations, it is 
straightforward to determine the components of the electromagnetic
tensor by means of
\begin{equation}
F_{\mu\nu} = A_{\nu,\mu} - A_{\mu,\nu}.
\label{omog}
\end{equation}
We explicitly have:
\begin{equation}
F_{\mu\nu} = \int{e^{i\,\bmit{k}\cdot\bmit{x}}\ 
\left [\beta^+(\bmit{k})\,f_{\mu\nu}^+\ e^{i(k x^0 + \psi^+)} +
\beta^-(\bmit{k})\,f_{\mu\nu}^-\ e^{-i(k x^0 - \psi^-)}\right ]
d^3k}
\label{Fmunu}
\end{equation}
where (dot means derivation with respect to $x^3 - x^0$):
\begin{eqnarray}
f_{03}^\pm &=& 
- i\ k_s\,\left(b^s - \frac{1}{2}\,h^s_{\ r}\,b^r\right)
\label{f03} \\
f_{0p}^\pm &=&
i\ \left[
\pm k\,\left(b^p + \frac{1}{2}\,h^p_{\ r}\,b^r\right) \mp 
b^p \frac{k_r k_s h^{rs}}{2\,\left(k \pm k_3 \right)}
\mp \frac{k_p\,k_s}{k\pm k_3}
\left(b^s - \frac{1}{2}\,h^s_{\ r}\,b^r\right)\right]
-\frac{1}{2} \dot h^p_{\ r}\,b^r
\label{f0p} \\
f_{12}^\pm &=&
i\ \left[
k_1\ \left(b^2 + \frac{1}{2}\,h^2_{\ r}\,b^r\right) -
k_2\ \left(b^1 + \frac{1}{2}\,h^1_{\ r}\,b^r\right)\right] 
\label{f12} \\
f_{p3}^\pm &=&
i\ \left[
- k_3\ \left(b^p + \frac{1}{2}\,h^p_{\ r}\,b^r\right) \mp 
b^p \frac{k_r k_s h^{rs}}{2\,\left(k \pm k_3 \right)}
\mp \frac{k_p\,k_s}{k\pm k_3}
\left(b^s - \frac{1}{2}\,h^s_{\ r}\,b^r\right)\right]
-\frac{1}{2} \dot h^p_{\ r}\,b^r
\label{fp3}
\end{eqnarray}

Maxwell equations in vacuum, $F^{\mu\nu}_{\ \ ;\nu}=0$, in the 
linear approximation of the chosen TT gauge reference frame [see 
Eq.~(\ref{lincond})] read:
\begin{equation}
\left(\eta^{\mu\alpha}\,\eta^{\nu\beta} - 
\eta^{\mu\alpha}\,h^{\nu\beta} -
\eta^{\nu\beta}\,h^{\mu\alpha}\right)\,F_{\alpha\beta,\nu} +
h^{\mu\alpha}_{\ \ ,\nu}\,\eta^{\nu\beta}\,F_{\alpha\beta} = 0,
\label{maxwell}
\end{equation}
where terms of the order of $h^2$ and $\chi\,h^2$ are negligible.
Actually the electromagnetic field components 
[Eqs.~(\ref{Fmunu})--(\ref{fp3})] satisfy Eq.~(\ref{maxwell}) with the 
same degree of precision. The homogeneous Maxwell equations are 
identically fulfilled because of Eq.~(\ref{omog}).

Therefore our solution solves the problem of the propagation of an 
electromagnetic field in a gravitational wave background with the only 
assumption of a small perturbation to a flat metric, thus generalizing 
the result of~\cite{mon98} to any gravitational wave frequency.

\section{Application}

As an example let us consider the case of an 
electromagnetic field bounced between two conducting parallel walls 
a distance $L$ apart. 
%We consider two cases in order to show the 
%peculiar features of the propagation of electromagnetic radiation in a 
%gravitational wave field. 
%
In the absence of gravitational waves, the physical features of the 
problem we are concerned with do not depend on the specific 
orientation of the apparatus. This is due to the isotropy of flat 
space--time. However this isotropy breaks down in the presence of a 
radiative gravitational field propagating along a given direction.
Because of this anisotropy, electromagnetic waves propagating in 
opposite directions are in general affected in different ways [see 
Eqs.~(\ref{psi}) and~(\ref{Fmunu})]. A consequence of this 
circumstance is the impossibility of obtaining stationary solutions.
Only if the propagation of the electromagnetic field lies on the 
plane perpendicular to the propagation of the gravitational wave, a 
stationary solution does exist. In fact, only in this case, the 
perturbation to the phase of the electromagnetic wave is the same for 
opposite directions [see Eqs.~(\ref{psi}) and~(\ref{Fmunu})]. The 
peculiarity of this plane also arises if we notice that two 
different points lying on it undergo the same gravitational 
perturbation. 

In Subsec. A we consider the last case and compare our result 
with the already existing solutions obtained by splitting of the 
electromagnetic field \cite{coo68,coo93} (see Sec. I). 
We find out that the latter 
approach lies in the framework of linearized theory only in the definite 
range of gravitational wave frequencies given by the relation:
\begin{equation}
%h \ll 
%\frac{k}{\chi} \ll \frac{1}{h}
k\,h \ll \chi \ll \frac{k}{h}
\label{splitrange}
\end{equation}
%when the 
%first inequality in Eq.~(\ref{condiz}) holds.
This is different from what was pointed out in Ref.~\cite{coo68}, 
where such a condition was mandatorily needed for the 
linearized theory to be valid. Indeed this is merely the 
condition for the validity of the splitting procedure.
We think that this circumstance is due to the expansion in power of
the gravitational wave amplitude of the electromagnetic phase, 
expansion that is needed by the splitting method. This kind of 
procedure (expansion of a phase) is not in general correct, as it is 
shown, for instance, by the long experience in perturbative methods 
in both classical and relativistic celestial mechanics (e.g. 
calculation of the advance of the perihelion of Mercury 
\cite{ber42}).
%
%From the mathematics point of view this is 
%due to the expansion in power of the gravitational wave amplitude of 
%the electromagnetic phase. 

In Subsec. B we show, in a particular case, how the non stationariness 
of the solution induces dissipative effects on the conducting planes.

\subsection{Comparison with the splitting procedure}

As a first case we assume the walls to be orthogonal to the 
$x^1$--axis, placed at $x^1=0$ and $x^1=L$, respectively.
Let us consider the particular solution to the de Rham equations that,
when $h_{\mu\nu}$ is switched off, 
describes a stationary electromagnetic field propagating along the 
$x^1$ direction with a linear polarization along the $x^2$--axis.
For the sake of simplicity we also consider a linearly 
polarized gravitational wave for which $H_\times = 0$ (see Sec. II).
This solution is: 
\begin{eqnarray}
F_{02}&=&\sum_{n} -2\,{\cal B}_n\,\sin{(k_n x^1)}
\left[\left(1 - H f\right)\,
\sin\left(k_n x^0 + \frac{H k_n}{2}\,f^{(-1)}\right) - 
\frac{H}{2\,k_n}\,\dot{f}\,\cos\left(k_n x^0 + 
\frac{H k_n}{2}\,f^{(-1)}\right)
\right]
\label{F02}\\
F_{23}&=&\sum_{n} H{\cal B}_n\,\sin{(k_n x^1)}\,\left[f \sin\left(
k_n x^0 + \frac{H k_n}{2}\,f^{(-1)}\right) + 
\frac{\dot{f}}{k_n} \cos\left(k_n x^0 + 
\frac{H k_n}{2}\,f^{(-1)}\right)\right] 
\label{F23}\\
F_{12}&=&\sum_{n} 2\,{\cal B}_n \left(1 - \frac{H}{2} f\right)\,
\cos{(k_n x^1)}\,\cos\left(k_n x^0 + \frac{Hk_n}{2} f^{(-1)}\right)
\label{F12}
\end{eqnarray}
where $k_n=\frac{n \pi}{L}$ and $n$ is an integer. 
We notice that the solution is still stationary in the $x^1$--direction. 
One can see that the gravitational wave induces a new component in the 
electromagnetic tensor (namely $F_{23}$).
For a direct comparison to Ref.~\cite{coo68} we set 
$f(x^3-x^0) = -\cos{\chi(x^3-x^0)}$, and
${\cal B}_n = {\cal B}_{\bar{n}}\,\delta_{n\bar{n}}$. Performing 
a formal expansion in $H$, the electromagnetic field 
components are found to be a sum of the flat space--time solution and a 
perturbation, that is:
\begin{equation}
F_{\mu\nu} =\, 
^{(0)}\!F_{\mu\nu}+\,
^{(1)}\!F_{\mu\nu}
\label{split}
\end{equation}
where:
\begin{eqnarray}
^{(0)}\!F_{02} &=& - 2\,{\cal B}_\nb\,\sin{(k_\nb \,x^1)}\,
\sin{(k_\nb \,x^0)}
\label{0split02}\\
^{(0)}\!F_{23} &=& 0
\label{0split23}\\ 
^{(0)}\!F_{12} &=&  2\,{\cal B}_\nb\,\cos{(k_\nb \,x^1)}\, 
\cos{(k_\nb \,x^0)}
\label{0split12}
\end{eqnarray}
are the ``unperturbed'' solutions, while
\begin{eqnarray}
^{(1)}\!F_{02} &=& -  2\,H\,{\cal B}_\nb\,\sin{(k_\nb \,x^1)}\,
\left[\cos{\chi (x^3-x^0)}\,\sin{(k_\nb \,x^1)} - \frac{1}{2}\,
\left(\frac{k_\nb}{\chi}+\frac{\chi}{k_\nb}\right)\,
\sin{\chi (x^3-x^0)}\,\cos{(k_\nb \,x^0)}\right]
\label{1split02}\\
^{(1)}\!F_{23} &=& -  H\,{\cal B}_\nb\,\sin{(k_\nb \,x^1)}\,
\left[\cos{\chi (x^3-x^0)}\,\sin{(k_\nb \,x^0)} - 
\frac{\chi}{k_\nb}\,\sin{\chi (z - t)}\,\cos{(k_\nb \,x^0)}\right]
\label{1split23}\\
^{(1)}\!F_{12} &=&  H\,{\cal B}_\nb\,\cos{(k_\nb \,x^1)}\,
\left[\cos{\chi (x^3-x^0)}\,\,\cos{(k_\nb \,x^1)} + 
\frac{k_\nb}{\chi}\,\sin{\chi (z - t)}\,\sin{(k_\nb \,x^0)}\right]
\label{1split12}
\end{eqnarray}
would represent the perturbation due to the gravitational wave. 
Actually from the above expressions one immediately achieves that 
the formal expansion of the exact solution is meaningless when 
condition ~(\ref{splitrange}) does not hold true. In fact, in this case 
the ``perturbation'' terms are even greater than the 
``unperturbed'' ones. 
On the other hand if Eq.~(\ref{splitrange}) does 
hold, the splitting approach~\cite{coo68,coo93} yields a result that
 coincides with the 
formal expansion in $H$ of our solution, but in a rotated 
reference frame.

It is important to notice  
that the solution expressed by 
Eqs.~(\ref{F02})--(\ref{F12}) can be cast in the form
\begin{equation}
A = \sum_{n} a_n(x^0,x^3-x^0)\,A_n(\bmit{x}),
\label{sumA}
\end{equation}
where $A$ stands for any electromagnetic tensor component, and 
$A_n$ are the same as in a flat space--time. The effect of the 
gravitational wave 
results in a modulation of the $a_n$ coefficients, and not in a 
creation of any new mode $A_{n'}$ ($n'\neq n$) even if 
$\chi=|k_n\pm k_{n'}|$.

\subsection{Dissipative effects}

As a second case of application we consider the walls to be orthogonal to 
the bisector to the $x^1$--$x^3$ plane, placed in $x^1=-x^3$ and 
$x^1=\sqrt{2}\,L -x^3$, respectively.
In a similar way to what we have done in the previous case, we 
consider a solution of the de Rham equation describing, when 
$h_{\mu\nu}$ vanishes, a stationary electromagnetic field propagating 
along the $x^1=x^3$ direction with a linear polarization along the 
$x^2$--axis. We set $h_\times=0$ again.
One has:
\begin{eqnarray}
F_{02} &=& \sum_{n}
-2\,{\cal B}_n \sin{\gm}\,\left[\left(1-H\,f\right) 
\sin{\alf} \right.
\nonumber \\
&-&\left.\frac{H}{2\,k_n}\,\dot f \cos{\alf}\right]  
\label{f02s} \\
&+& \frac{{\cal B}_n H f}{\sqrt{2}}\,\cos{\gm}\,\cos{\alf} 
\nonumber \\
F_{23} &=& \sum_{n}
-\sqrt{2}\,{\cal B}_n \left(1-H\,f\right)\,\cos{\gm}\,\cos{\alf} 
\label{f23s} \\
&+& {\cal B}_n H\,\sin{\gm} \left[f\,\sin{\alf} + \frac{\dot f}{k_n}
\,\cos{\alf}\right] \nonumber \\
F_{12} &=& \sum_{n}
\sqrt{2}\,{\cal B}_n\,\left(1-\frac{H\,f}{2}\right)\,
\cos{\gm}\,\cos{\alf}.
\label{f12s}
\end{eqnarray}
As in the previous case, the gravitational wave does not 
create any new modes of oscillation $A_{n'}$ [see Eq.~(\ref{sumA})].

The most interesting feature of the solution is that the 
parallel component to the surface of the electric field does not 
vanish on the walls any more. 
This results in a {\em gravitationally induced} ohmic power loss 
within the conductors which is 
to be added to usual losses (such as those due to thermal vibrations 
of the conducting planes) already existing in flat space--time and not 
depending on the orientation of the apparatus. 
Consequently if the former kind of loss is at least comparable to the 
latter one it would be possible, in principle, to infer the 
presence of a gravitational wave by measuring the time decay of a 
resonant cavity.

We now proceed to estimate the {\em gravitationally induced} power 
loss per unit surface when
\begin{equation}
\frac{H\,k_n}{\chi} \gg H.
\label{acoop}
\end{equation}
This assumption leads to interesting results.
In order to study dissipative currents induced by the gravitational 
wave on the walls, we only need to retain the biggest part of the 
field; it is noticeable that 
%this one is of order zero in $H$, 
all terms proportional to $H$ are negligible
when Eq.~(\ref{acoop}) is fulfilled. One gets 
(${\cal B}_n= \delta_{n\nb}\,{\cal B}_\nb$):
\begin{eqnarray}
F_{02}&=&-2 {\cal B}_{\bar n}\,\sin{\left(\Psi-
\frac{k_{\bar n} H f^{(-1)}}{2\,\sqrt{2}}\right)}
\,\sin{\left(k_{\bar n} x^0+
\frac{k_{\bar n} H f^{(-1)}}{2}\right)}
\label{f02d}\\
F_{23}&=&-\sqrt{2} {\cal B}_{\bar n}\,\cos{\left(
\Psi-\frac{k_{\bar n} H f^{(-1)}}{2\,\sqrt{2}}\right)}
\,\cos{\left(k_{\bar n} x^0+
\frac{k_{\bar n} H f^{(-1)}}{2}\right)}
\label{f23d}\\
F_{12}&=&-F_{23}
\label{f12d}
\end{eqnarray}
where $\Psi = 0$ if $x^1 = -x^3$ and $\Psi = k_{\bar n}\,L$ 
when $x^1 = -x^3+\sqrt{2}\,L$.
If we set ${\cal U}=\frac{dU}{da}$, where $U$ is the electromagnetic 
energy and $da$ stands for the surface element parallel to the walls, 
then the ohmic loss is given by~\cite{j75}:
\begin{equation}
\frac{d{\cal U}}{dt} = 2\ \frac{c^2\,\bmit{E}_\|^2}{8\pi\,\delta\, \omega} 
\label{loss}
\end{equation}
where $c$ is the speed of light (in the following we drop the 
convention $c=1$, hence $x^0 = c t$ and $\Omega = \chi\,c$ stands for the 
gravitational wave frequency), $\bmit{E}_\|$ is the electric field 
component parallel to the surface, $\delta$ is the skin depth,
and $\omega=c\,k_\nb$ the electromagnetic frequency; 
factor $2$ takes into account the fact that there are two walls.  
Therefore the equation describing the rate of energy loss per unit 
surface is:
\begin{equation}
-\,\frac{1}{{\cal U}}\,\frac{d{\cal U}}{dt} = 
\frac{2\,c^2}{L\delta\,\omega}\,\sin^2{\left(
\frac{k_{\bar n}\,H\,f^{(-1)}}{2 \sqrt{2}}\right)},
\label{eqloss}
\end{equation}
where an average over $\frac{2\pi}{\omega}$ has been 
performed. The solution to the above equation can be easily obtained 
assuming $\chi\,x^3 \sim 0$:
\begin{equation}
{\cal U}(t) = {\cal U}_0\,\exp{\left[-\int_0^t 
\frac{2\,c^2}{L\delta\,\omega}\,\sin^2{\left(
\frac{k_{\bar n}\,H\,f^{(-1)}(-t')}{2 \sqrt{2}}\right)}\,dt'
\right].}
\label{solloss}
\end{equation}
If we set $f(x) = \cos{(\chi x)}$ and make the further assumption 
(whose validity must be checked at the end) $\Omega\,t \ll 1$, 
Eq.~(\ref{solloss}) becomes:
\begin{equation}
{\cal U}(t) = {\cal U}_0\,\exp{\left[
-\,\frac{2}{3}\,\frac{c^2}{L \delta\,\omega}\,
\left(\frac{k_{\bar n}\,H}{2 \sqrt{2}\,\chi}\right)^2\,
\Omega^2\,t^3\right]}  
\label{sollossapp}
\end{equation}
The time after which the energy falls by a factor $1/e$ (time decay) 
is:
\begin{equation}                        
\tau = \left[ \frac{3}{2}\,\frac{L \delta\,\omega}{\Omega^2\,c^2}\,
\left(\frac{2\,\sqrt{2}\,\Omega}{\omega H}\right)^2
\right]^{\frac{1}{3}}.
\label{tau}
\end{equation}
If the gravitational wave source considered is a binary pulsar as, for 
instance the PSR $1913+16$~\cite{tay75}, then we find:
\begin{equation}
\tau = 5\ \left(\delta\,L\right)^{\frac{1}{3}}\ 10^4\ sec \simeq
50\ sec \ll \frac{1}{\Omega} \sim 5\ 10^3\ sec
\label{PSR}
\end{equation}
where we have set $\omega = 10^{15}\ rad\,sec^{-1}$ and 
$\left(\delta\,L\right)^{\frac{1}{3}} \simeq 10^{-3}\,cm^{2/3}$ 
(e.g.~\cite{rossi}). Unfortunately this time lies many order of 
magnitude above the usual --- non {\em gravitationally 
induced} --- optical resonant cavity decay times.

However if we consider a coalescing compact binary~\cite{coc96} we get:
\begin{equation}
\Omega\,\tau - \frac{\sin{(2\,\Omega\,\tau)}}{2} = 
\frac{8\,\delta\,L\,\Omega^3}{c^2\,\omega\,H^2}
\end{equation}
where Eq.~(\ref{solloss}) has been used. Analysis of this equation, 
by taking $m_1 \sim m_2 \sim 3\ m_\odot$, 
shows that the shortest decay time is obtained when the time to 
coalescing ranges within $2\ 10^6\ sec \sim 23\ d$. In this case 
$\tau \sim 0.8\ sec$, to be compared with the decay times 
of the order $1$--$10\ msec$ in a cavity. 
Therefore we see that this time is long, as compared to 
presently--measured decay times, but not so long to rule out the 
possibility to measure it in the future.

\section{Conclusion}

We have found the exact solution to the linearized Maxwell equations 
for a gravitational background described by a plane gravitational 
wave. We have not made any assumption over the shape or 
frequency of the gravitational wave. The only approximation used is
the linearized Einstein theory, within which Eqs.~(\ref{maxwell}) hold. 
%
%The knowledge of the four--vector potential enables to 
%determine the 
%electromagnetic field components, which are the only physical measurable 
%quantities.

As an application we have evaluated the electromagnetic field bounced
between two conducting walls. A comparison with known approximated 
results has been performed. We have found that, if 
Eq.~(\ref{splitrange}) is met, the formal expansion in $H$ of our 
solution reduces to the approximated one.
Besides we have found out that the gravitational wave does not cause any 
new mode $A_{n'}$ [see Eq.~(\ref{sumA})] to appear in the cavity. 
Furthermore we have evaluated a
possible measurable effect of the interaction with low frequency 
gravitational waves. We have shown that in general an ohmic loss on 
the walls is produced; this results in a decrease of the stored 
electromagnetic energy.

\acknowledgments

The authors are pleased to thank V. Guidi for reading of the 
manuscript.

\end{document}